\documentstyle[11pt,emulateapj5]{aastex}
\newcommand{\CIV}{\hbox{{\rm C}\kern 0.1em{\sc iv}}}
\newcommand{\MgII}{\hbox{{\rm Mg}\kern 0.1em{\sc ii}}}
\newcommand{\HI}{\hbox{{\rm H}\kern 0.1em{\sc i}}}
\newcommand{\HII}{\hbox{{\rm H}\kern 0.1em{\sc ii}}}
\newcommand{\Ly}{\hbox{{\rm Ly}\kern 0.1em$\alpha$}}
\newcommand{\lya}{\hbox{{\rm Ly}\kern 0.1em$\alpha$}}
\newcommand{\Ha}{\hbox{{\rm H}\kern 0.1em$\alpha$}}

\newcommand{\kms}{km~s$^{-1}$}

\newcommand{\compl}{24.35}	%24.42 	%completeness at 50% for an aperture of 2 FWHM
\newcommand{\complAB}{24.80}	%24.88	%completeness at 50% in AB mag for an aperture of 2FWHM
\newcommand{\Nexp}{\hbox{$N_{\rm exp}$}}
\newcommand{\Nobs}{\hbox{$N_{\rm obs}$}}
\newcommand{\Pdla}{\hbox{$P_{\rm DLA}$}}
\newcommand{\Ppdla}{\hbox{$P_{+}$}}
\newcommand{\Pmdla}{\hbox{$P_{-}$}}
\newcommand{\Imin}{22.5}
\newcommand{\IminAB}{23.0}
\newcommand{\nobs}{7}
\newcommand{\nexpected}{2.70}

\slugcomment{accepted to ApJ}

\begin{document}

\title{Clustering of Galaxies at $z\sim 3$ around the probable Damped Ly-alpha Absorber towards QSO APM 08279+5255
    }%end title

\author{Nicolas Bouch\'e~\altaffilmark{1,2}  }
\affil{Dept. of Astronomy,  University of Massachusetts-Amherst,  Amherst, MA 01003 USA; bouche@astro.umass.edu}
\author{James D. Lowenthal~\altaffilmark{1}}
\affil{ Dept. of Astronomy, Smith College, Northampton, MA 01063 USA; james@earth.ast.smith.edu}
\altaffiltext{1}{Visiting Astronomer, Kitt Peak National Observatory, National Optical
    Astronomy Observatory, which is operated by the Association of
    Universities for Research in Astronomy, Inc. (AURA) under cooperative
    agreement with the National Science Foundation.}
\altaffiltext{2}{EARA Marie-Curie Fellow at the Institute of Astronomy, Cambridge, UK}

%%%%%%%%%%%%%%%%%%%%   ABSTRACT  %%%%%%%%%%%%%%%%%%%%%%%
\begin{abstract}
We present results on the clustering of Lyman break galaxies (LBGs) around  
 a probable damped \lya\ absorption (DLA) line cloud at $z_{\rm abs}=2.974$ from
  deep $UBVI$ images of the field 
containing the quasar APM 08279+5255 ($z_{\rm em}=3.91$).
The large area covered by our images, 0.31  $\deg^2$ or $\sim40\times 40$~Mpc co-moving 
 at redshift $z=3$, and their depth, $\mu_{I,\rm AB}(sky) \simeq 27.6$~mag~arcsec$^{-2}$,
 allow us to identify $\sim 450$  LBG candidates  brighter than $I_{AB}=\complAB$
 at $2.75<z<3.25$  both close (50~kpc) to the DLA  line-of-sight and 
up to 20 Mpc (co-moving) from the DLA, i.e. physically unrelated.
LBG candidates were identified using photometric redshift techniques
that include the   $I$ magnitude as a prior estimate in addition to the colors.
 The two are combined using Bayes' theorem. 
This helps to break the degenerancies that occur in a pure spectral template fitting scheme. 
The overall rms is $\sigma_z\simeq 0.15$ at $z\sim 3$  based on our analysis of photometric redshifts in the HDF-N.

From the redshift likelihood distributions, we selected LBG galaxies 
within a redshift slice of width $W_z=0.15(\simeq\sigma_z)$ centered on the redshift of the  DLA $z_{\rm abs}$. 

Within that redshift slice, we find an enhancement of galaxies near the DLA using
both the surface density  ($\Sigma/\Sigma_g \simeq 3$) and an  estimator of the 3-D spatial overdensity
($n/\overline n_g \sim 5\pm 3$).
The surface overdensity $\Sigma/\Sigma_g$ is significant at the $>95$\%\ significance level 
on scales  $2.5<r_\theta<5$~Mpc co-moving.   The overdensity can not be
related to the QSO environment since the QSO is at $z_{em}=3.91$. 
These results imply that some DLA could reside in high density regions.  

We search within 45\arcsec\ from the line of sight for galaxies responsible for the DLA, and
find one candidate with $z_{\rm phot}=3.03$ that is
26\arcsec\ (145~kpc physical) away. From its magnitude $I=24.65\pm0.2$,
its luminosity is $M_{I,\rm AB}=-21.35$. Due to its large impact parameter, however,
this galaxy is not a likely candidate for the absorber.

\end{abstract}

\keywords{galaxies: evolution --- galaxies: high-redshift ---
 quasars: absorption lines --- quasars: individual (APM08279+5255)}

%%%%%%%%%%%%%%%%%%%%%%%%% INTRODUCTION %%%%%%%%%%%%%%%%%%%%%%%
\section{Introduction}

Damped \Ly\ absorbers (DLAs) give rise
to the largest neutral hydrogen (\HI) column densities ($N_{\HI}>2\times 10^{20}$~cm$^{-2}$)
of quasar absorption line systems.
 They are important because they contain the
largest reservoir  of \HI\ at high redshift, and
their detailed nature 
have been the subject of an ongoing debate.
 \citet{wolfe86} argued that DLAs are massive gas rich disks    based on the
similarity between the amount of \HI\ contained in DLAs and the
stellar masses of disk galaxies today.  Later, \citet{wolfe95} and
\citet{prochaska97} argued that DLA kinematics
 are consistent with those expected from lines of
sight intercepting massive rotating  thick gaseous disks.

However,  theoretical simulations of galaxy formation show  that a
large range of structures  and morphologies
can give rise to DLAs
\citep[e.g.][]{katz96,haehnelt98,mcdonald99}.   DLAs would arise 
from the combined effect of a massive central galaxy and/or a number of
smaller satellites in a virialized halo \citep{maller00} or filaments
\citep[e.g.][]{haehnelt98}, i.e.,   DLAs would lie in  
over-dense regions.  

Several approaches have been used to investigate the
nature of high$-z$ DLAs.  Broad band imaging surveys of DLAs have been used
since \citet{steidel92a}; \citet{steidel92b}  were  unsuccessful  in finding the absorber,
 but did find  large numbers of Lyman break galaxies (LBGs) \citep[e.g.][]{steidel93}.
In addition,  it has become clear that  DLAs   (1) show  little
evolution from $z = 4$ to $z = 1$ in their contribution to the matter density of the Universe 
 $\Omega_{\rm DLA}(z) $  \citep[e.g.][and references therein]{rao00,boissier03};
(2) show little metallicity evolution \citep{pettini99,prochaska02} from $z = 3$ to $z = 0.3$;
(3) have heterogeneous chemical properties \citep{pettini00};
(4) have various morphologies \citep{lebrun97,rao00}; and
(5) have low star formation rates \citep{kulkarni00,bouche01}.

These results indicate that DLAs  do not participate significantly
in the overall chemical enrichment of the Universe and hence do not trace star formation
in any straightforward way. 
Both these observations and theoretical work show that DLAs  are
not  a single type of galaxy, but  are merely characteristic of a  
type of region: namely, one with a large neutral column density \citep[e.g.][]{khersonsky96}.

Similarly to broad-band imaging, early  attempts  to detect
diffuse \Ly\ emission from DLAs at $z>2$ using
 deep 
  narrow band imaging \citep[e.g.][]{lowenthal95} 
   did not reveal the absorber, but unveiled a few companion \Ly\ emitters,
hinting at the clustering of galaxies around DLAs. 
This prompted \citet{wolfe93a} to calculate the two-point correlation function
at $z=2.6$ and to conclude that indeed \Ly\ emitters are clustered near DLAs.
Some recent \Ly\ searches succeed in unveiling the absorber \citep[e.g.][]{fynbo99}.

Here, we seek to constrain the   matter distribution around  DLAs at $z\sim 3$  
using LBGs as tracers of the large scale structure.
Specifically,  are the DLAs in {\it over- or under-dense} regions? 
Previous attempts at answering this question include the    investigation of \citet{gawiser01}, 
which  ruled out any clustering between  galaxies and  a DLA at $z\sim 4$.
Recently, \citet{adelberger03}  found only two galaxies
within 265\arcsec ($\sim 5.7h^{-1}$~Mpc) and within $|\Delta z|<0.0125$ ($\sim 8h^{-1}$~Mpc) of four  DLAs
at $z\sim 3$ whereas 5.96 were expected. They conclude from the same null hypothesis test used here (see below) that
the lack of galaxies near the four DLAs is significant at $>90$\%\ level and argue
that this is evidence that DLAs and LBGs 'do not reside in the same parts
of the universe'.
  
In contrast, \cite{francis93} reported the discovery of super-clustering of sub-DLAs at $z\sim 2.3$.
 Recent   narrow-band imaging (Fynbo et al. 2003, private communication)
of the same field shows that spectroscopically confirmed \Ly-emitting galaxies are clustered at the
redshift of the two strongest \HI\  clouds.
These observations and others \citep[e.g.][]{dodorico02} show that at least
some DLAs reside in overdense regions.

The field of   QSO 3C336   shows an over-density of
galaxies around an intermediate redshift DLA at $z=0.656$ \citep{steidel92a,steidel97,bouche01}.  
 \citet{steidel92a} suggested  the possibility of a cluster at $z=0.656$.
From a sample of spectroscopically confirmed   galaxies within 50\arcsec\ (0.4 $h^{-1}$ Mpc co-moving) of the QSO line of sight, 
\citet{steidel97} showed that galaxies are  clustered in redshift around   the metal-line systems
  \citep[see][Fig~7]{steidel97}. A closer analysis reveals that they are even more clustered 
around  the DLA: the number of galaxies per unit redshift,
 $\mathrm dN/\mathrm dz$, peaks at $z=0.656$ (aside from the peak at $z_{em}=0.927$). 
 Indeed, out of   11 galaxies at $0.4<z<0.85$,  four galaxies lie within $\Delta_z<0.03$ 
 (9,000 \kms) of the DLA on scales of $\sim 200h^{-1}$ Mpc (co-moving).

 In this paper, we present results of our survey of LBGs around a DLA  at $z=2.974$  
towards QSO APM 08279+5255.  This field is the
first of four. The results on the other three (one with some spectroscopy) 
are still pending and will be presented in \citet{bouche03} along with the details of our
procedure and data analysis.
Our study is similar to the investigations of \citet{gawiser01} and \citet{adelberger03}; 
  however, our MOSAIC sky coverage is
more than 25 times larger, providing us with both extended spatial
sensitivity and a built-in control sample of the background surface
density of LBGs.

Throughout this paper, we adopt $\Omega_M=0.3$, $\Omega_\Lambda=0.7$
and $H_o=100 h$~\kms~Mpc$^{-1}$; thus, at $z\sim3$, $1\arcsec$
corresponds to  $\sim 21.5h^{-1}$~kpc and
$1\arcmin$ to $\sim 1.29h^{-1}$~Mpc, both  {\em co-moving}. At that redshift,
$H(z)\sim 4.46 H_o$, so $\delta z=0.1$ corresponds to  
$67 h^{-1}$~Mpc in co-moving coordinates.
Magnitudes are Vega based unless stated otherwise.

%%%%%%%%%%%%%%%%%%%%%%%%%%%%   DATA     %%%%%%%%%%%%%%%%%%%%%
\section{The data } %Observations and Data Reduction}
\subsection{The APM 08279+5255 field}

This field, like the other three in our survey, was selected for the presence
of a DLA at $z\sim 3$ and with requirement $z_{abs}\ll z_{em}$. The redshift
range $z\sim3$ is ideal for selecting LBGs efficiently using standard photometric filters.

Even taking into account its gravitational lens magnification  
$10\; -100\times$,  this APM 08279+5255 ranks among  the most luminous objects
 in the Universe \citep[e.g.][]{irwin98,lewis99}.
The DLA at $z_{abs}=2.974$ was first reported by \citet{petitjean00} based
on a high resolution spectrum (6 \kms) of the QSO APM 08279+5255 ($z_{\rm em}=3.91$),
obtained by \citet{ellison99}~\footnote{Their study was aimed at investigating
\Ly\ forest between $3.109<z<3.701$.} at the
10-meter Keck telescope with   HIRES.  
Due to uncertainty in the continuum,
the DLA's column density is poorly constrained: $19.8<\log N(\HI) <20.3$.
The absorber could therefore technically be a `sub-DLA'. 
However,  the   \Ly\   equivalent width  $W_{\rm rest}(\Ly )>4.8$\AA\ \citep{petitjean00}  
and its strong \MgII\ equivalent width  $W_{\rm rest}(\MgII)>0.6$\AA\ \citep{koba02} 
supports its classification as a damped cloud.

\citet{petitjean00} estimated a minimum cloud size   $r_{\rm DLA} >200 h_{75}^{-1}$ pc from the
image separation of the QSO.
  The most robust  estimate  is from \citet{pana03}  
  who found $r_{\rm DLA}>450h^{-1}$~pc (assuming the gravitational lens  is at $z\sim 1$).
  The QSO and DLA properties are summarized in Table~\ref{table:field}.

\subsection{Observations, data reduction \& completeness}
The observations, summarized in Table~\ref{table:summary}, were carried
out with the MOSAIC camera \citep{jacoby98} at the Kitt Peak
National Observatory  4-m telescope on  UT February 7th and 8th, 2000. The wide field
imager MOSAIC has eight 2k$\times$4k thinned SITe CCDs. With 0.258\arcsec\
per pixel, it has a field of view of 36 arcmin on a side.
 
Our field was imaged through the $UBV\&I$ filters  
with   integration times as shown in Table~\ref{table:summary}.  The four
nights were photometric.
A detailed description of the data reduction and calibration of our  survey
is deferred to \citet{bouche03}.  To summarize, the images were
bias-subtracted, corrected for the MOSAIC ghost pupil image, dome- and sky-flatfielded, cleaned for
cosmic-rays, deprojected and registered onto a common frame, before
being co-added. 
All magnitudes quoted below are measured in a $2\times FWHM$ 
diameter aperture, where the FWHM is the seeing.  
The resulting magnitude limits are shown in Table~\ref{table:summary}.

From Monte-Carlo simulations, we estimated   our completeness as a function of magnitude, and is
shown in the inset of Fig.~\ref{fig:sample} as the dotted line. Our 50\% completeness level is  $m_I=\compl$.
Using the transformation to convert our $I$-band  magnitudes into AB magnitudes, i.e. 
   $I_{\rm AB}=m_I+0.47$, this  corresponds to $I_{\rm AB}=24.8$ ($\mathcal R\simeq 25$) and
   to $\sim 0.6 L^*$, where we used $m^*_{\mathcal R}=24.5$ at $z\simeq 3$ \citep{steidel99}.
In terms of $L^*$(today), it corresponds to $L/L^*\simeq 3$
  using  a  distance modulus $m-M=46.25-5\log h$ at $z=3$   in our cosmology,
 $M^*_{g,{\rm AB}}-5\log h=-20.04$  from SDSS \citep{blanton01},
 and  a $k-$correction  between our $I$ and the SDSS $g$ filter of $- 0.30$
  \citep[assuming the Irr SED of][]{coleman80}.
 
%%%%%%%%%%%%%%%%%%%%%%%%   Selection %%%%%%%%%%%%%%%%%%%%%%
\subsection{Sample selection }

\begin{figure} 
\plotone{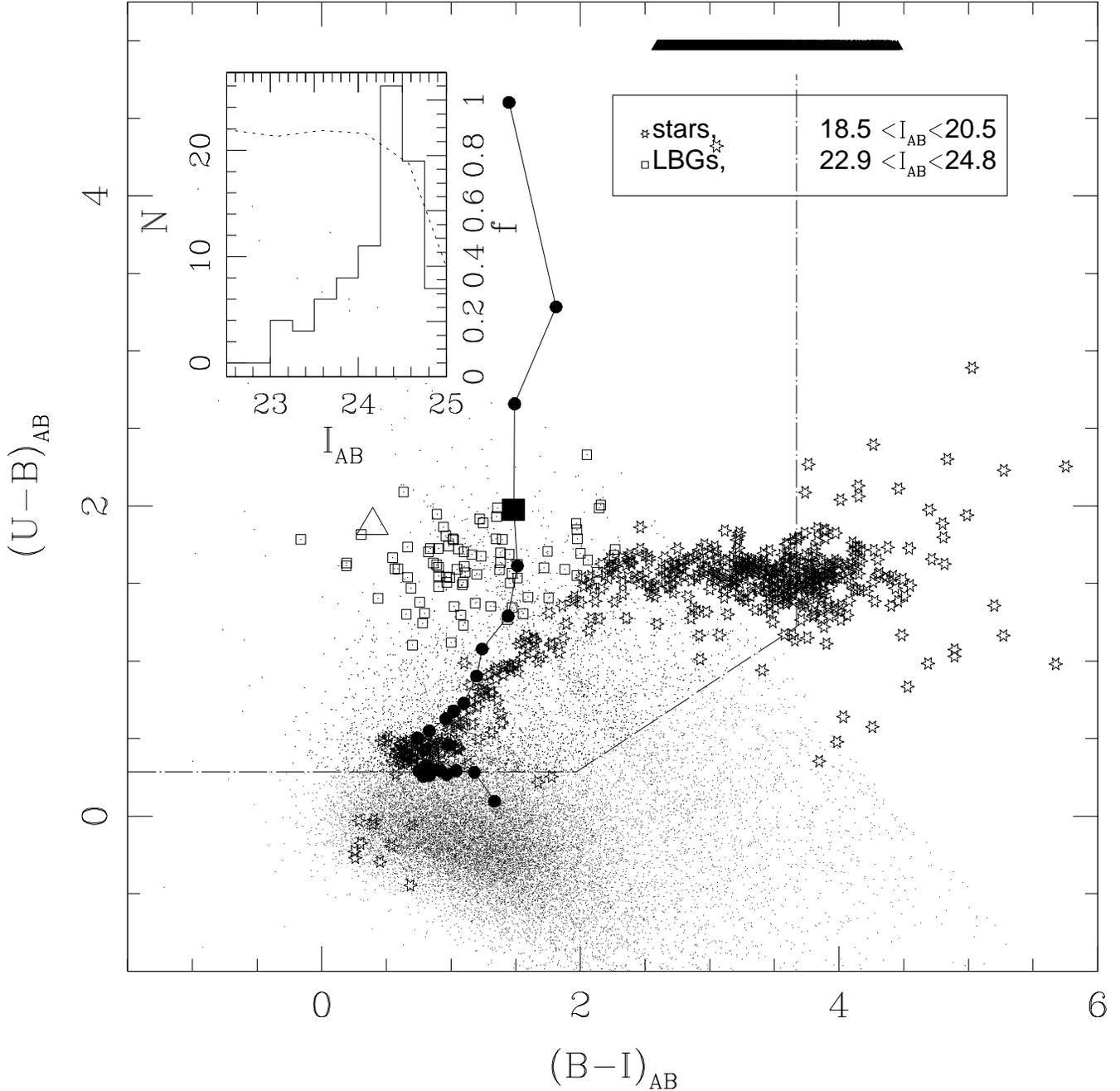} 
\caption{Color-color $(U-B)_{AB}$, $(B-I)_{AB}$ diagram. Each dot represents one of our $\sim30,000$ objects in our image
with $\IminAB<I_{AB}<\complAB$. Open squares are our LBGs candidates in the redshift slice centered on the DLA ($z_{abs}=2.974$),
 i.e. with $\Pdla>0.5$ (see text). 
Objects with $U$ and $B$ upper limits (filled triangles)  have their $U-B$ set to an arbitrary constant ($=5$),
and the $B$ magnitude limit is used in $B-I$. 
Stars show the locus of stellar objects with $18.5<I_{AB}<20.5$.  
The open triangle is the closest LBG to the  line of sight, which has an impact parameter of 145~kpc (see \S\ discussion). 
The  evolutionary
track of an Irr SED (averaged over different dust content, from $A_v=0$ to $A_v=1.2$) is shown in redshift steps of 0.1
for illustrative purposes; the $z=3$
 mark is shown with the large filled square.   We pre-selected
objects within the color cut  shown with the dot-dashed lines.
 The inset shows their   number counts ($N$)  as a function of magnitude ($I_{AB}$). The  dotted line shows our 
completeness ($f$).
 \label{fig:sample} }
\end{figure}

Sources were detected in the $I$-band using SExtractor \citep{bertin96},
and aperture magnitudes were measured in all four bands using
identical matched apertures defined in the $I$-band image to extract 30,000 objects with $I>\Imin$. 
Fig.~\ref{fig:sample} shows the position of each object in the $(U-B),(B-I)$ color space.
We first selected a subset (1/3) of our  $I$-band catalog using  the   color cuts
shown by the dot-dashed lines in Fig.~\ref{fig:sample}. 
This removes most of the $z\leq1$ objects and reduces the amount of computing in the next step.
We then estimated  photometric redshifts of all objects in our subsample using
the template fitting routine {\it Hyperz} from \citet{bolzonella00}~\footnote{
We note that we used the MOSAIC filter curves and the CCD response curves in performing the SED fits.
}. This constrains the redshift mainly via the shape of the SED at $\lambda_{rest}=912$\AA.
In addition, we  used the apparent magnitude  as a redshift prior following the Bayesian prescription of \citet{benitez00}.
The advantage of this last step is that   it breaks most
of the degenerancies of the SED fitting procedure (i.e. 4000\AA\ break taken as a \Ly\ break) as shown by
 \citet{benitez00}. 
Finally, from our catalog now complemented with photometric redshifts $z_{\rm phot}$ and likelihood distributions $P(z)$,
 we selected galaxies with   $\Imin<I<\compl$  
 and computed the   probability of each to be in three redshift slices of width $W_z=0.15$: 
 (1)   $\Pdla \equiv 	P(z_{\rm DLA}\pm 0.075)$, 
 (2)	$\Ppdla \equiv 	P([z_{\rm DLA}+0.15]\pm 0.075)$,
 (3) $\Pmdla \equiv 	P([z_{\rm DLA}-0.15]\pm 0.075)$.
 
Out of 428 LBG candidates   at redshift $2.75<z<3.25$,   
    84 objects with $\Pdla>0.5$  constitute  our sample of candidates
  in a redshift slice centered on the DLA and are shown in Fig.~\ref{fig:sample} as open squares.
    This 0.5 threshold maximizes the number of objects  
   and minimizes the likely outliers: $P_{outlier}\equiv P_z[|\Delta_z|>0.2(1+z)]\leq 0.1$, following \citet{benitez00}.  
   As a posteriori check, the distribution of LBG candidates in Fig.~\ref{fig:sample} shows   that we are
   not excluding significant $z\sim 3$ objects in our color pre-selection. The number counts
   are shown in the inset of Fig.~\ref{fig:sample}.
Similarly 89 (17) LBG  have $\Ppdla>0.5$  ($\Pmdla>0.5$), respectively.

The slice width ($W_z=0.15$) was chosen to  correspond to the rms of the residuals between
photometric and spectroscopic redshifts $\sigma(\Delta_z)$.
 $\sigma(\Delta_z)$ is  $ \simeq 0.15$ ($\sigma (\Delta z/(1+z)) \simeq 0.06$) 
in the redshift range $2.8<z<3.5$  from our analysis of photometric redshifts in the HDF-N. 
The residuals are similar to the accuracy in other studies \citep[e.g.][]{fernandez-soto01}.  
We defer   the  detailed description of our photometric redshift analysis  to \citet{bouche03}. 

The  slice width adopted here produces the largest sample in the smallest redshift slice. Our
results are not affected when we chose a slice of   half or twice  the current size. 
At $z=3$, the width  $W_z=0.15$ corresponds to $\sim100$~Mpc (co-moving) or 25~Mpc (physical) assuming $h=0.75$.
Peculiar velocities of 7,500\kms\ would be required to have an effect on our results. This figure
is much larger than any peculiar velocity in local clusters    of galaxies. At $z=3$, peculiar
velocities are likely to be smaller than 1,000\kms. Therefore, peculiar velocities are not
affecting our results.

%%%%%%%%%%%%%%%%%%%%%%%%% RESULTS  %%%%%%%%%%%%%%%%%%%%%%%%%%%%%%
\section{Results}

\subsection{Clustering analysis}
\label{section:clusterana}

 Hierarchical models of galaxy formation predict that    massive galaxies will   likely be found
in regions of  high density, whereas low-mass galaxies are more uniformly distributed. 
 This produces  an enhancement of the clustering of high-mass galaxies  
since they tend to form near each other in rare overdense regions. 
Therefore, clustering  is a probe of  the mass distribution of galaxies. For instance,
if LBGs cluster around  DLAs, it would indicate that  DLAs reside in massive  dark halos.

A natural tool  to investigate the clustering of galaxies is the two-point auto-correlation function $\xi_{gg}(r)=(r/r_o)^{-\gamma}$.
However, our redshift accuracy $\sigma(\Delta_z)\simeq0.15$ corresponds to $100h^{-1}$~Mpc.
Therefore, without more accurate redshifts,  the angular auto-correlation function $w_{gg}(\theta)$ is a more
relevant statistic.
Similarly to the two-point  angular correlation function $w_{gg}$,  
  the cross-correlation
between DLAs and LBGs, $\omega_{dg}$,  gives the excess probability of
finding an LBG in a small annulus of area $d\Omega$ between $\theta$ and
$\theta+d\theta$ from a known DLA. The expected number of LBGs in that annulus is then:
 \begin{equation}
 N(LBG|DLA)=\overline{\Sigma}_g(1+\omega_{dg}(\theta)) d\Omega, \label{crosseq}
\end{equation}
where $\overline{\Sigma}_g$  is the underlying
surface density of LBGs which is  $\overline\Sigma_g=0.041h^2$~Mpc$^{-2}$ in our slice centered on the DLA.
 Thanks to the large field
of view of the MOSAIC imager, there is little ambiguity in estimating the background surface density of LBGs.
The background density is measured at $15<r<20h^{-1}$~Mpc, i.e. where galaxies 
 are physically unrelated to the DLA.

Estimating $w_{dg}$ accurately would require a large number of DLAs.
Therefore, we investigated the spatial distribution of LBG candidates  in
two limiting cases: (i) assuming the LBG-DLA cross-correlation is the same as the LBG auto-correlation
and (ii) assuming the cross-correlation is zero.  In addition, we used 
a 3-D estimator of the cross-correlation. In the next two sections, we present our results
and discussion of this analysis.

\subsection{Clustering results}
\label{section:clustering}

\begin{figure} 
\plotone{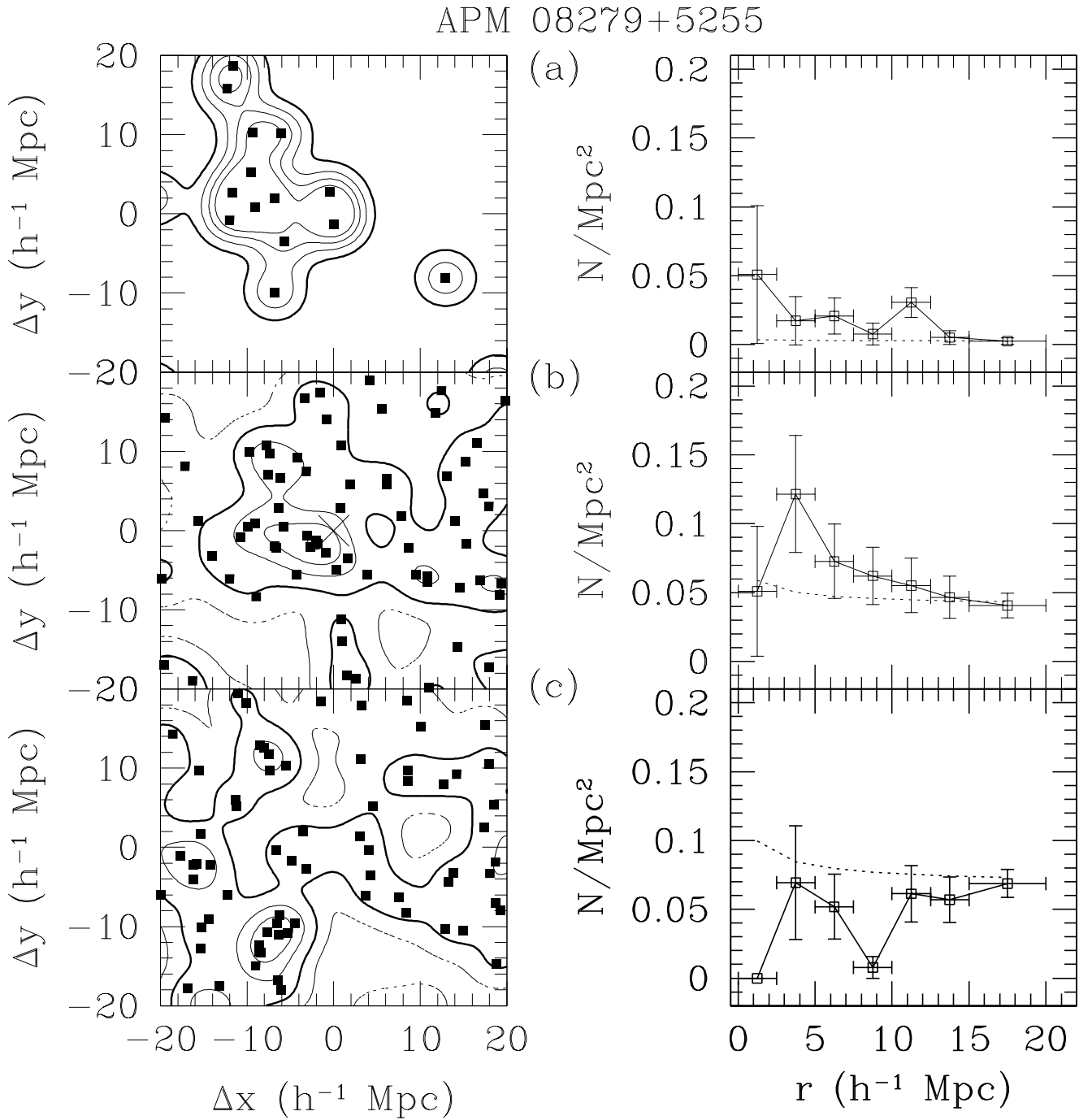} 
\caption{The spatial distribution of LBG
candidates  (left) and the radial surface distribution (right)
for three different redshift slices:
a)   slice centered at $z=z_{DLA}+0.15$, with $\Ppdla >0.5$;
b)    slice centered at $z=z_{DLA} $, with $\Pdla >0.5$; and
c)    slice centered at $z=z_{DLA}-0.15$, with $\Pmdla >0.5$.
Panel b) is centered on the DLA, marked by the cross.
Panels a) and c) shows the distribution of galaxies unrelated to the DLA.
Contours on the left panels shows the surface densities $\delta=\Sigma/\overline \Sigma-1$, where $\overline \Sigma$
is the mean density over the entire field. 
The thick contour shows $\delta=0$; the continous contours
show $\delta=1,2,3$; the dotted countours show $\delta=-1$. %  1.5,  2.5, 3.5, 4.5$h^2$~Mpc$^{-2}$.
The error bars on the right panels are computed using bootstrap resampling. 
The QSO line-of-sight is maked with a cross.
For galaxies  within the slice centered on the DLA (b),
     the surface density distribution  shows
a  peak (significant at $>95$\%; see text) at $\sim 4$~Mpc from  the  DLA.
  This  indicates an
over-density of  galaxies by a factor $\sim 3 \times $  on scales $2.5-5$~Mpc from the DLA
compared to the averaged background  galaxy density  $\overline{\Sigma}_g$. %_g=<\Sigma>$.
 \label{fig:clustering}
 }
\end{figure}

Figure~\ref{fig:clustering} (left panels) shows the spatial distribution of LBG
candidates in the three redshift slices, each of width $W_z=0.15$:
a) the slice at $z=z_{DLA}-0.15$ with $\Pmdla >0.5$,
b) the slice centered on the DLA at $z_{\rm DLA}=2.974$,
 and c) the slice at $z=z_{\rm DLA}+0.15$ with $\Ppdla >0.5$.
We smoothed the surface  density with a Gaussian kernel of 2.5$h^{-1}$~Mpc, shown 
with the contours in Fig.~\ref{fig:clustering}. The contours show
the surface  density $\delta=(\Sigma-<\Sigma>)/<\Sigma>$, where
$<\Sigma>$ is the average surface density of the  field.  The thick contour shows $\delta=0$,
the continuous contours show  $\delta= 1,2,3$, and the dotted contour shows $\delta=-1$.

 The right  panels of Fig.~\ref{fig:clustering} 
 show the radial cumulative surface density centered on the DLA
 for the same three slices.
The error bars are derived using bootstrap resampling method \citet{mo92}.

The middle panels of Fig.~\ref{fig:clustering} show that 
the surface density of LBG candidates is higher   by a factor $\sim 3 \times $    within $2.5-5$~Mpc
than the averaged background  galaxy density  $\overline{\Sigma}_g$.  
This shows that at least some DLAs could reside in overdense regions.
and implies a possible   cross-correlation between this DLA and LBGs.

\begin{figure} 
\plotone{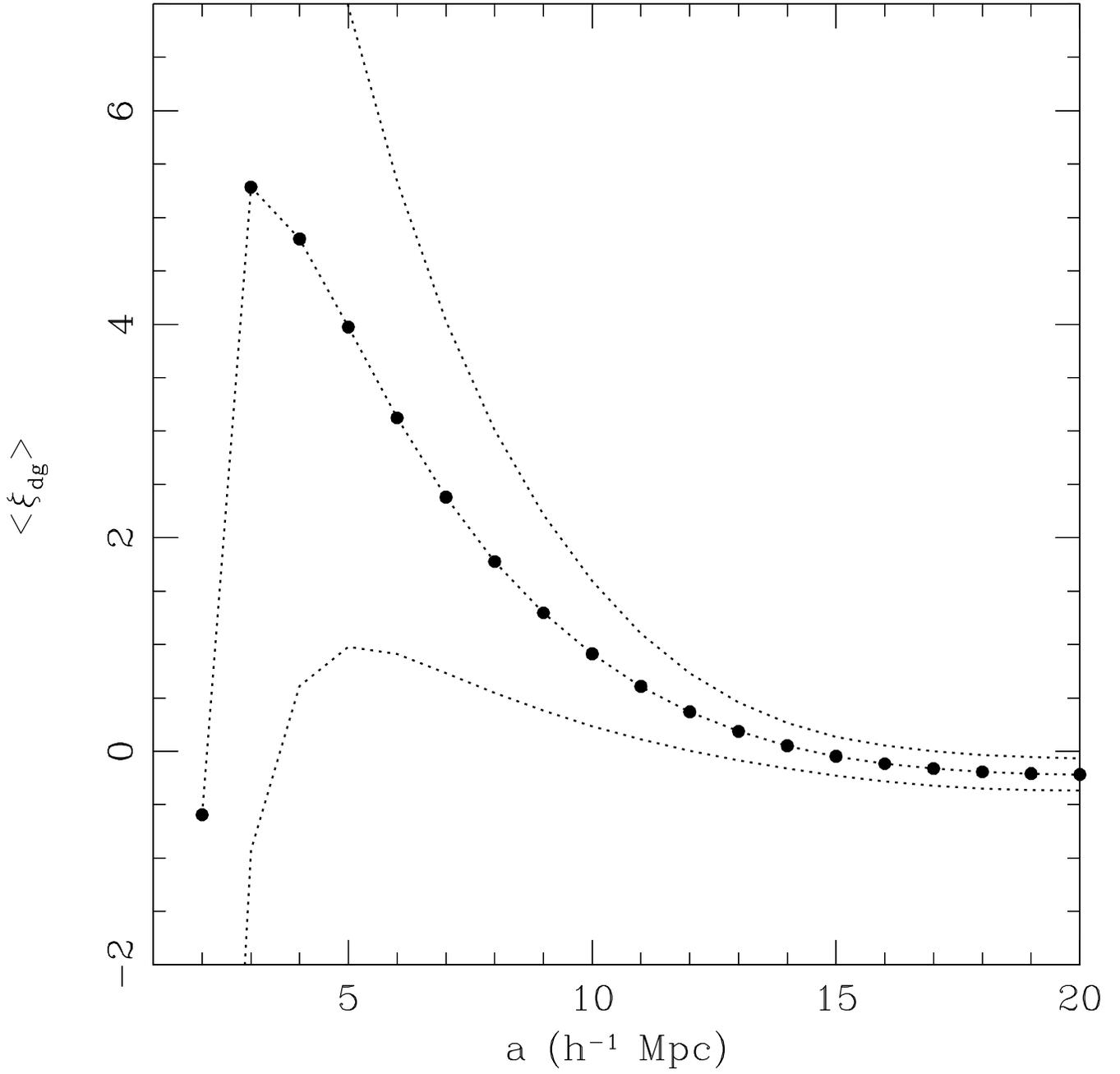} 
\caption{The volume averaged (using a Gaussian
window function) of the cross-correlation function $\xi_{dg}$ , following \citet{eisenstein02}
is shown as a function of the scale length $a$ of the window function. The ($1\sigma$) errors,
 shown as the dotted lines, include the Poission variance and the variance from the clustering
 of the galaxies.
\label{fig:eisenstein}
}
\end{figure}

For the slice centered on the DLA, Fig.~\ref{fig:clustering}(b),	
we used  the estimator of the cross-correlation proposed by \citet{eisenstein02},	
which gives the volume average of the cross-correlation $<\overline \xi_{dg}>$.
 This estimator uses weightings that produce a spherically
symmetric aperture rather than the  cylindrical one used above.
 Fig.~\ref{fig:eisenstein} shows $<\overline \xi_{dg}>$ as a function
of the scale length of the Gaussian window.
We find a 3-D overdensity  $1+<\overline \xi_{dg}>=n/\overline n_g \sim 4.97 (\pm 2.99)$ using a Gaussian
window of 5$h^{-1}$~Mpc. 
The errors  include the shot noise and the contribution of the clustering of the galaxies.
We again conclude that this DLA seems to reside in an overdense region.
 
 Next, in order to assess the significance of our result, we examined the
  two limiting situations presented  
   and some possible  sources  of  contamination.

In the first case, where we assume the LBG-DLA cross-correlation is the same as the LBG auto-correlation,
i.e. $\xi_{dg}=\xi_{gg}$, we test whether  this   excess of 
galaxies around the absorber could be caused by the LBG clustering.
A qualitative way to answer this  is given
by the dotted line in Fig~\ref{fig:clustering}, which shows the expected number  of
galaxies around the DLA  
 due  to the clustering of LBGs. That number  is naturally found from
Eq.~\ref{crosseq} integrated  from $0$ to $r_\theta^{max}$
using the known auto-correlation of LBGs, $\xi_{gg}$, from \citet{adelberger03}
converted into $\omega(\theta)$ using our redshift distribution.
 Clearly the angular clustering of LBGs
alone cannot explain our data.

 A more quantitative  answer is given by computing the expected
number of galaxies \Nexp$(r)$ again assuming $\xi_{dg}=\xi_{gg}$, 
and comparing it with the observed number of galaxies \Nobs. 
  Table~\ref{table:stat} shows the values for at various radii. E.g., within $2.5<r_\theta<5 h^{-1}$~Mpc,
  we find \nobs\ galaxies  while \nexpected\ were expected.
Could \Nobs\ and \Nexp\
be drawn from the same Poisson distribution of true mean  \Nexp? 
The answer is given by testing the null hypothesis 
``\Nobs\ and \Nexp\ are drawn from the same Poisson distribution''
 versus   ``\Nobs\ is greater than \Nexp'',
 i.e. the observed overdensity is  statistically significant.
At $2.5<r_\theta < 5 h^{-1}$~Mpc, we find there is a probability of finding N$_{\rm obs}\geq \nobs$ galaxies 
(under the null hypothesis with  \Nexp$=\nexpected$)
of $<1$\% ($=$P-value shown in Table~\ref{table:stat}). This implies
the null hypothesis ought to be rejected, and 
the observed number of galaxies  
is {\it larger than} what one would expect from the LBG auto-correlation
at the $>99$\%\ significance level. 

In the second   limiting case $\omega_{dg}=0$, we test whether
a purely random fluctuation in $\Sigma$ could cause our result.
 This would decrease the expected
number of galaxies slightly and therefore increase the significance level
 (see Table~\ref{table:stat}).
 
  \subsection{Discussion}
\label{section:discussion}

Our results are not sensitive to the background density.  The background density,
itself, is a strong function of the magnitude limit and of the redshift slice width.
A change of the background density $\Sigma_g$ will change significantly both the observed
and expected number of galaxies throughout the field. However, both
are affected in the same way from  Eq.~\ref{crosseq} by $\Sigma_g$.
The effect on the overdensity cancels out,
since the overdensity is the excess of observed  to the expected number of galaxies.
The same relative overdensity by a factor of $\sim 3\times$ is observed with different
magnitude cuts from 24 to 24.5, and with a width half or twice our current redshift width.

Alternative causes for the apparent excess of galaxies include  contamination from low redshift sources (stars or
galaxies) and  signal from the   cross-correlation between LBGs and  other quasar
 absorption line clouds besides the DLA.

   We estimate our contamination from  low-redshift
galaxies to be $\sim$ 10\% at $2.8<z<3.5$ based  
on  our photometric redshift  analysis in the HDF \citep{bouche03}.
Another source of contamination is from faint stars with color similar to those of our LBG candidates.
To  estimate the stellar contamination,
 we used our star counts up to $m_I\leq23.5$ and extrapolate  them to our magnitude limit,
  assuming the counts to be constant at $m_I>23.5$.
   This gives an upper limit on our stellar contamination of $<7$\%. 
We note that both of these contaminations would  dilute any clustering signal.

A high-resolution spectrum of
this QSO \citep{ellison99} shows there are 12 \CIV\ metal-line
systems with 3.1$<z<$3.5 and 4 with 2.97$<z<$3.10 \citep{pana02}. 
However, the cross-correlation between \CIV\ systems and LBGs is less strong ($r_o=3h^{-1}$~Mpc) than
the LBG auto-correlation \citep[$r_o=4h^{-1}$~Mpc,][]{adelberger03}.
 Therefore, on scales $5h^{-1}$~Mpc,
the cross-correlation between  \CIV\ systems and LBGs cannot 
bias our results, for it would have to be larger than the 
LBG auto-correlation.

We close this section with a discussion of our search for the DLA itself.
The QSO is not only lensed, but it also saturated the detector, thus
preventing us from any attempt at PSF subtraction. Due to the brightness of the QSO,
we are  not sensitive  to faint objects with impact parameters less than 5--6\arcsec.
From our  source catalog, we searched for all objects within 45\arcsec\ (285~kpc in physical units) 
that have colors, i.e. photometric redshift, consistent with being at $z\sim 3$.
We find one object at an impact parameter of $r_\theta=145$~kpc (physical)
that did not make it into our final catalogs for it is  fainter than our completeness limit by 0.3mag.
It has a magnitude of $24.65\pm0.2$ ($M_{I,\rm AB}=-21.35$) and a redshift $z_{\rm phot}=3.03$.
  The colors of this galaxy are shown in Fig.~\ref{fig:sample} by the open triangle.
Besides this galaxy, the closest object that did end up in our final catalog is 2.35~Mpc (co-moving)
or 590~kpc (physical) away. We regard the galaxy 145~kpc away as an unlikely, but possible, candidate for the DLA.

\section{Summary and Conclusions}

To summarize, in the field of APM 08279+5255, we find
\nobs\ galaxies with $r_\theta <4h^{-1}$~Mpc (186\arcsec) and within
$|\Delta_z|=0.075$, whereas one would expect \nexpected\ assuming
that the cross-correlation is given by the LBG auto-correlation, i.e.
that DLAs would be galaxies similar to LBGs.
The hypothesis that this is due to  either the galaxy autocorrelation
 or to random fluctuation is rejected at
the $>95$\% level. 
We estimated our contamination from stars or low-$z$ galaxies to be $<17$\%, since 
such contamination would only dilute our cross-correlation, so our results gives
 a lower limit on the cross-correlation.

We therefore conclude that,
compared to the averaged background  galaxy density  $\overline{\Sigma}_g$,
 there is  an overdensity $\Sigma/\Sigma_g\simeq 3$ of LBGs within $2.5-5$~Mpc around this  $z_{\rm abs}=2.974$ DLA.
 This scale   corresponds to
only $1.3$~Mpc in physical coordinates at $z=3$, when clusters would
still be in formation and have core radii larger than 8~Mpc.
Large scale structures at high redshifts shouldn't be unexpected.  Recently, \citet{shima03} 
 reported the discovery of  large-scale structure of Lyman alpha emitters   at $z=4.86$,
 12$h_{70}^{-1}$ Mpc (co-moving) in radius, 
  based on wide-field imaging with Suprime-Cam on the Subaru telescope.

Our result  implies a possible correlation between this DLA and LBGs, and 
that at least some DLAs reside in overdense regions. However,
preliminary results in two of our other fields \citep{bouche03} indicate 
  a weak and an anti-correlation between LBGs and the other two DLAs. 
  This could imply that not all DLAs reside in the same environment.  
 This reinforces the idea that 
DLAs do not  likely represent a single type of object, but are a wide variety of
objects with high neutral hydrogen column density.

As a final note, 
the galaxy with the smallest impact parameter in our final  slice centered on the DLA    is   $  2.35h^{-1}$~Mpc (co-moving)
or 590~kpc (physical) away from the QSO line of sight. In addition,
we search for all objects within 45\arcsec (285~kpc physical) and
find an LBG candidate with $z_{\rm phot}=3.03$ at $r_\theta=145$~kpc, with a luminosity of 
$M_{I,\rm AB}=-21.35$.  Due to the large impact parameter, this galaxy is
unlikely to be responsible for the DLA, although it may be associated with it.
 
 We are currently obtaining   multi-object spectra  with  Gemini+GMOS 
 of our LBG candidates within   $5~h^{-1}$~Mpc of the QSO line-of-sight
to investigate further the clustering of LBGs and the DLA.

\acknowledgments NB acknowledges the support from the EARA Marie Curie
 Fellowship under contract HPMT-CT-2000-00132 and from the European
 Training Network `The Physics of the Intergalactic Medium'.  JDL
 acknowledges support from NSF grant AST-0206016. 
  We thank M. Haehnelt, T. Theuns and   J. Schaye for  useful comments on an earlier draft.
  We thank A. Bunker and E. Gawiser   for a careful reading of  the manuscript, which improved
the quality of the paper.
  Thanks to the Institute of Astronomy, where part of this work was  accomplished.

\clearpage
\begin{deluxetable}{ll}
\tablewidth{0pt}
\tablecaption{Properties of the field.\label{table:field} }
\tablehead{
	\colhead{Parameter} & \colhead{Value} 
	}
\startdata
	\multicolumn{2}{c}{QSO APM 08279+5255 ($z_{\rm em}=3.91$) } \\ 
	\hline \\
	R.A. (J2000) 	\dotfill 	&  $08^h31^m41.6^s$ \\
	Decl. (J2000) 	\dotfill 	&    $52\arcdeg 45\arcmin 17\arcsec$\\
	$m_R$	(mag)   	\dotfill	&	$15.2$ \\
	$A_U$\tablenotemark{a} (mag) \dotfill	&	0.20 \\
	$E_{B-V}$\tablenotemark{a} (mag)\dotfill	& 0.04\\
	\hfill \\
	\multicolumn{2}{c}{Damped Ly-$\alpha$  Cloud ($z_{\rm abs}=2.974$) } \\
	\hline \\
	$\log N_{HI}$\tablenotemark{b}	\dotfill & $19.8-20.3$ \\
	$[Fe/H]$\tablenotemark{b} 	\dotfill & -2.31 \\
	
\enddata
\tablerefs{(a) from \citet{schlegel98}, averaged over the field; (b) \citet{petitjean00} } 
\end{deluxetable}

	%%%%%%%%%% Observations  %%%%%%%%%%%%%%%%
	%%%%%%%%%%%%%%%%%%%%%%%%%%%%%%%%%%%%%%%
	
\begin{deluxetable}{ccccccc}
\tablecaption{Summary of observations.\label{table:summary} }
\tablewidth{0pt}
\tabletypesize{\small}
\tablehead{
	\colhead{} &
	\colhead{Exp. /Frames} &
	\colhead{Airmass} &
	\colhead{FWHM} &
	\colhead{$\mu_{I} (1\sigma$)  	} &
	\colhead{m$_{I}(3\sigma$)  \tablenotemark{1}	} &
	\colhead{Completeness 50\% \tablenotemark{1}} \\
	\colhead{Filter} &
	\colhead{(sec./\#)} &
	\colhead{(min-max)} &
	\colhead{(arcsec)} &
	\colhead{(m$_{\rm AB}/\mbox{arcsec}^2$)} &
	\colhead{(m$_{\rm AB}$)} &
	%\colhead{(m$_{\rm AB}/(2\times FWHM)^2$)} &
	\colhead{(m$_{\rm AB}$)}
}
\startdata
%FIELD		    &   Exp/#   &      X    &   PSF    &   SB\(1sig)  &     m_lim(3sig) &  Completeness
 	    	  $U$ & 13500/15  & 1.07-1.17 & 1.1 &28.61 & 26.68& \\
		 $ B$ &  2100/7   & 1.14-1.27 & 1.1 &28.37 & 26.42& \\
		  $V$ &  3000/10  & 1.21-1.37 & 1.2 &28.23 & 26.19&  \\
		  $I$ &  7590/20  & 1.11-1.62 & 1.1 &27.62 & 25.68&   \complAB \\ 
\enddata

\tablenotetext{1}{measured inside a  $2\times$ FWHM diameter aperture.}
\end{deluxetable}

%%%%%%%%%%%%%%%%%%%	Hypothesis Test

\begin{deluxetable}{lccccc}
\tablewidth{0pt}
\tablecaption{Null Hypothesis Test at various radii.\label{table:stat} }
\tablehead{
	\colhead{$r_\theta$ ($h^{-1}$~Mpc)} &  
	\colhead{$\Sigma$ ($h^2$~Mpc$^{-2}$)} 	&
	\colhead{\Nobs\tablenotemark{a}  } & 
	\colhead{\Nexp\tablenotemark{b}	}   & \colhead{P-value\tablenotemark{c}}
	 }
\startdata
 $0-2.5$\dotfill	&  0.05	& 1	& 1.04	        	& 0.27 (0.19) \\
 $2.5-5$\dotfill  	& 0.12	&\nobs\ 	& \nexpected\         	& $<$0.01 ($<$0.01) \\
 $5-10 $\dotfill  	& 0.07	 & 15   & 9.90			& 0.10 (0.07) \\ 
 $15-20$\dotfill  	& 0.04	& 16     & 16.71		& 0.50 (0.43)
\enddata
\tablenotetext{a}{Observed  number of neighbors.}
\tablenotetext{b}{Expected number of neighbors assuming $\xi_{dg}=\xi_{gg}$.}
\tablenotetext{c}{P-values of the null hypothesis, i.e. probability that  \Nobs$ > $\Nexp\  if \Nobs\ were drawn
from a distribution of true mean equal to  \Nexp, assuming  $\xi_{dg}=\xi_{gg}$ (assuming  $\omega_{dg}=0$).}
\end{deluxetable}
\clearpage

%%%UCP%%%
%\newpage
%\plotone{f1.eps}
%\newpage
%\plotone{f2.eps}
%\newpage
%\plotone{f3.eps}

\end{document}